\documentclass[aps,pre,twocolumn,groupedaddress,showpacs,floatfix]{revtex4}
\usepackage{amsmath}
\usepackage{amssymb}
\usepackage{graphicx}
\usepackage{epic}
\usepackage{eepic}
\usepackage{subfigure}
\begin{document}

\title{On the Importance of Hubs in Hopfield Complex Neuronal Networks under
  Attack} 
\author{E. P. Rodrigues}
\author{M. S. Barbosa}
\email[]{marconi@if.sc.usp.br}
\author{L. da F. Costa} 
\affiliation{Institute of Physics of S\~ao Carlos - University of S\~ao Paulo,
  S\~ao Carlos, SP, PO Box 369, 13560-970, Brazil.}
\date{\today}


\begin{abstract}
  The organizational principles behind the connectivity of a complex network
  are known to influence its behavior. In this work we investigate, using the
  Hopfield model, the influence of the network architecture on the performance
  for associative recall while the network is under hub and edge attack.
  We show, by using four different attack strategies, that although the
  importance of hubs is more definite for Barab\'asi-Albert neuronal
  networks, the random removal of the same amount of edges as in a
  hub may imply a greater reduction of memory recall.
\end{abstract}

\maketitle

\section{Introduction}

A great deal of the current interest on networks, neuronal or complex, stems
from the fact that the functionality of network systems is highly influenced,
and even determined, by their respective connectivity
(e.g.~\cite{review:2005}).  While most neuronal network models were initially
fully connected, increasing interest has been focused more recently on systems
with diluted (e.g.~\cite{forrest:1989,kurten:1990,miljkovic:2001,burcu:2004})
and/or structured
(e.g.~\cite{Stauffer:2003,costastauffer:2003,jun:2005,bianconi:2003,kim:2004,castillo:2004,torres:2004,torres:2005})
connectivity.  A series of interesting points and questions are implied by
such investigations, especially regarding structured networks of connections,
i.e. models departing from uniformly random connections.  For instance, given
that metabolic and spatial biological pressures constraining the number of
dendrites and axons in a neuronal cell, it is virtually impossible to have
fully-connected biological networks (e.g.~\cite{Karbowski:2001}).  What are
the functional implications of such unavoidable partial connections?  Which
partially connected structure (e.g.  random, small world, scale free) allows
the best functionalities?  Are hubs particularly important for scale-free
neuronal networks?  In which sense they affect the recall of patterns in
associative networks such as the Hopfield model?

Because they concentrate connections, hubs have been found to play a
fundamental role in defining the connectivity of scale free complex networks
(e.g.~\cite{Albert_Barab:2002,Newman:2003,survey:2005}).  For instance, as they
connected to many nodes, hubs provide bypasses between a large number of pair
of nodes, contributing to the overall reduction of the mean shortest path in
the network.  It is intuitive to expect that hubs would also be particularly
important for the dynamics of systems with scale-free connectivity, such as
the Hopfield complex neuronal networks considered in
~\cite{Stauffer:2003,costastauffer:2003}.  Such a question provided the main
motivation for the present article.  In order to try to answer it, we
performed node and edge attack on random and Hopfield complex networks and
monitor the effect over the recall performance.  Among the reported
experiments, we successively remove hubs from scale free Hopfield models, and
then remove the same number of edges by node and edge attack on random models.
Although hubs are confirmed to have special importance for the memory capacity
of the network, it was remarkably found that their removal is not so impacting
as the random elimination of the same number of edges from complex or random
networks.  One explanation for such a phenomenon is the fact that a hub is
most directly and strongly associated to a single bit of the patterns to be
trained and recovered.

The article starts by reviewing the Hopfield model and complex networks, and
follows by describing the simulation framework and the obtained results.  It
concludes by interpreting and discussing the results and their implications
for further researches.

\section{Methodology} The Hopfield model~\cite{hopfield:1982} is a model of
autoassociative memory which has been studied and generalized thoroughly since
its first appearance.  Today, the value of the original model is mostly
epistemologic, as practical implementations are constrained by several issues
such as limited efficiency for storing several patterns. Nevertheless, being a
simple and intuitive model, it provides an effective framework for performance
studies under varying conditions while allowing the quantification of the
impact of the choice of specific influencing elements.

In this paper we follow closely the implementation described
by~\cite{haykin:1999}. A set of $M$ binary string patterns are represented by
vectors as $P_i=[p_1,p_2,...,p_n]^T$, whose $N$ elements are in either of the
possible states $p_j=\pm 1$, randomly chosen. This information is stored in a
weight matrix $W$, according to the follow expression
\begin{equation}
W=\frac{1}{N}\sum_{i=1}^{M} P_i P_i^T -MI,
\end{equation}
In order to retrieve a pattern stored initially, say
$Q_n=[q_1,q_2,...,q_n]^T$, it is necessary to iterate the following expression
\begin{equation}\label{eq:recall}
Q_{n+1}=sgn \left[WQ_n\right],
\end{equation}
where $sng\left[. \right]$ is a hard limiting function. In this
implementation, whenever the value of the state of a neuron becomes exactly
zero, the previous state of that neuron is used instead, so that the neurons
are always either firing (+1) or quiescent (-1).  The element $w_{ij}$ of $W$
is the weight of the connection of neuron $i$ with neuron $j$ with the $W$
matrix representing a fully connected network, except for loops (i.e.
$w_{ii}=0$).

A diluted version of the Hopfield model still retaining pattern recognition
properties, see~\cite{Stauffer:2003}, can be obtained by
multiplying, in elementwise fashion, the weight matrix by a sparse
\emph{weight} connection matrix, build upon some assumed heuristics.

We consider in this paper two models of complex networks, namely the random
and Barab\'asi-Albert networks, as connection matrices underlying Hopfield
models. A network with $N$ nodes can be represented by a symmetric adjacency
matrix $C$ of rank $N$, with each element $c_{ij}$ assuming unit value for
connected nodes and null values for disconnected ones. We set $c_{ii}=0$ to
prevent self connections. The random network is obtained simply by choosing
connections $c_{ij}=1$ with probability $\gamma$. The parameter $\gamma$
therefore determines the density of connections.

In the Barab\'asi-Albert
model~\cite{Albert_Barab:2002,Barabasi_Ravasz:2001}, we start with a
small random network $\lambda$ with $m_0$ nodes. For each node $a_i$
the probability $p_i$ of a new connection is given as
$k(a_i)/\sum_{i}(k(a_i))$, where $k(a_i)$ is the node degree. Consider
a new node $b$ with $m$ edges.  A node $a_i$ of the initial network
$\lambda$ is allowed to connect to one of the $m$ edges of the new
node $b$ with the probability $p_i$.  In other words, the degree of
node $a_i$ defines its chances of receiving new connections,
suggesting the paradigm 'the rich gets richer'.

\section{Simulations}
\begin{figure*}
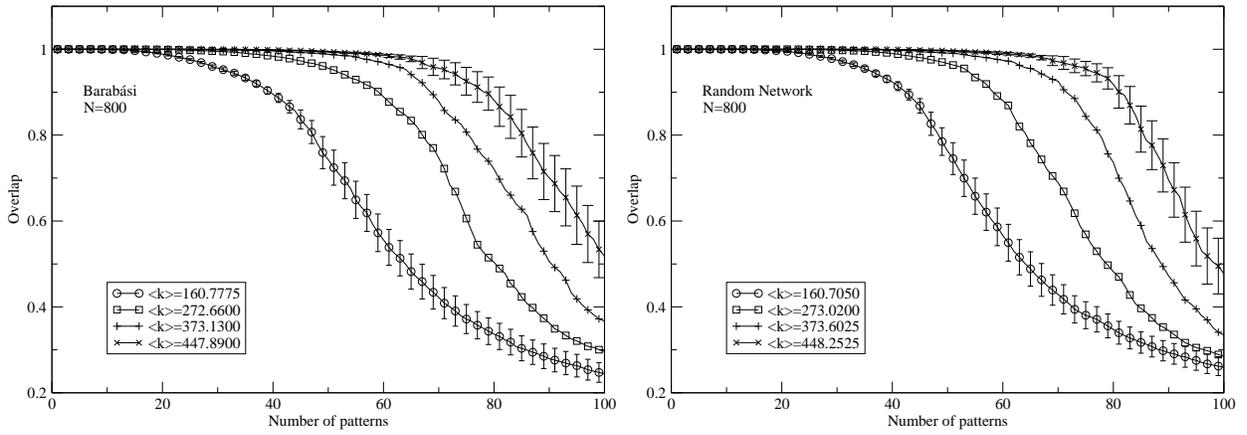

\includegraphics[scale=0.35,clip]{figura1_I.eps}
\includegraphics[scale=0.35,clip]{figura1_II.eps}
\caption{Overlap as a function of the number of patterns for Barab\'asi-Albert and Random
  Network with different $\left\langle k \right\rangle$.}
\label{diff_mean_k}
\end{figure*}

In this section we consider both kinds of network architectures (i.e. random
and scale free), characterized by their respective connection matrix $C$, as
defined in the previous section.  The training matrix $W$ of the Hopfield
model, calculated for a initial set of random patterns $P_n$, is diluted by
direct multiplication with the connection matrix $C$ of the network under
analysis. The pattern $P_1$ will stand for the reference pattern for these
experiments. We add $10\%$ of noise to $P_1$ before trying to recover $P_1$
through the iterative recall process, see Equation~\ref{eq:recall}. After
fifty interactions, an overlap index is calculated, measuring how much of the
recovered pattern $P_r$ matches the original pattern $P_1$. This process is
repeated for an increasing number of patterns $P_i$, $i=1,2,...,M$.  The
calculation of the overlap index is performed by
\begin{equation}
Overlap(i)=\frac{\{P_r \cdot P_1\}(i)}{N},
\end{equation}
where $N$ is the total number of neurons in the network and $i=1,...,M$. This
procedure produces an overlap curve as a function of the number o patterns
stored $i$, quantifying the degradation of performance as the number of
patterns grows. Figure~\ref{diff_mean_k} shows the overlap curves for
several experiments, each averaged over a hundred realizations, exploring both
Barab\'asi-Albert and Random network architectures with varying $\left\langle
  k\right\rangle $, implemented by changing $m$.

\begin{figure*}
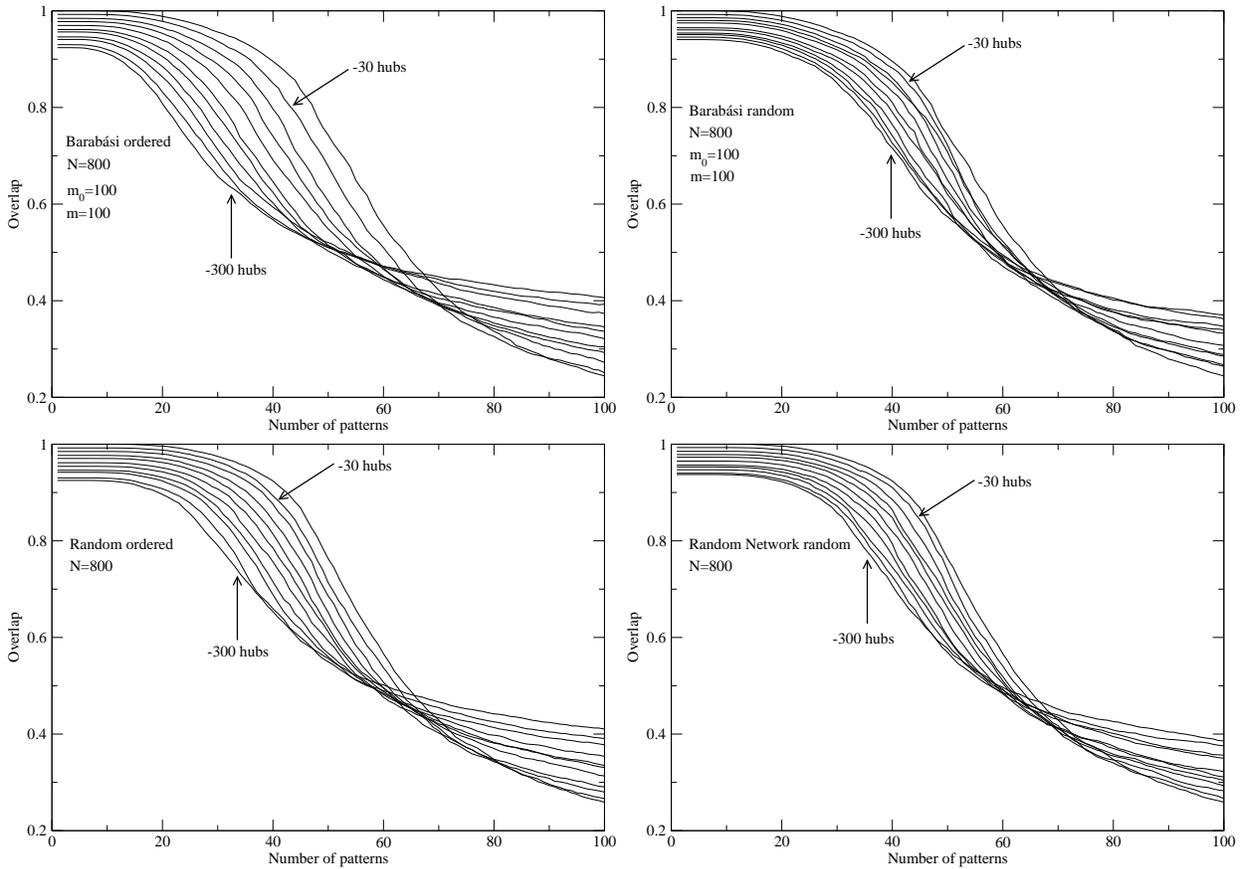

  \includegraphics[scale=0.35,clip]{figura2_I}
  \includegraphics[scale=0.35,clip]{figura2_II.eps}\\
  \includegraphics[scale=0.35,clip]{figura2_III.eps}
  \includegraphics[scale=0.35,clip]{figura2_IV.eps}
  \caption{Overlap as a function of the number of patterns for
    Barab\'asi-Albert and Random networks under both random and orderly hub
    attack .}
  \label{kill_hub}
\end{figure*}

To implement hub attack, in each experiment one hub is identified and
eliminated from $W$ by setting to zero all the elements of the connection
matrix along the respective column and row. We eliminate hubs either
sequentially, i.e., from the more connected to the less connected hub, or at
random, with uniform probability. Figure~\ref{kill_hub} shows four experiments
in which $30$ hubs, belonging to two types of networks (random or
Barb\'asi-Albert), were successively attacked in two manners (orderly or not).
The Barab\'asi-Albert networks were grown with parameters $m_0=m=100$. The
random network counterparts were obtained with equivalent $\left\langle
  k\right\rangle$. Both types of networks included $N=800$ neurons. As
expected, the overall recall performance decreased with the number of trained
patterns and as the hubs were eliminated.  The mean node degree $\left\langle
  k\right\rangle$ was also directly affected.  Marked differences appears then
in these cases, discriminating the impact of hub withdraws, orderly or not,
for each architecture. Note, for instance, that the Barabasi networks are more
sensitive to orderly hub removal than a random network with equivalent degree.
Note also that both networks respond essentially in the same manner to random
hub deletion.

\begin{figure*}
\centering
\includegraphics[scale=0.4,clip]{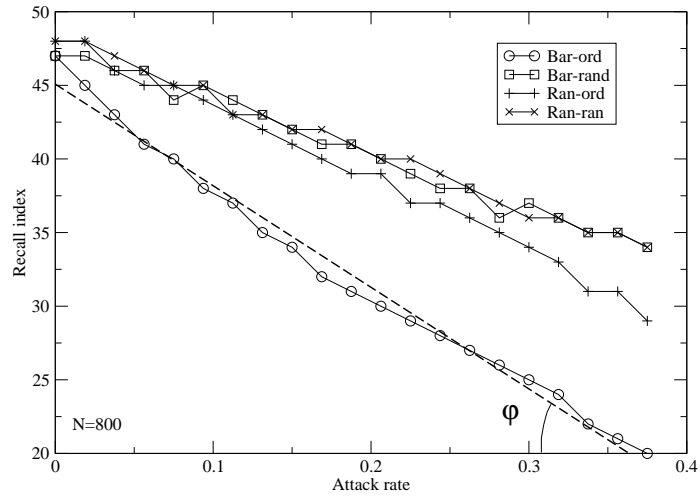}
\caption{Recall index as a function of the attack rate for $N=800$ and $\left\langle k
\right\rangle=160.7$}
\label{Rec_attack}
\end{figure*}

\begin{figure*}
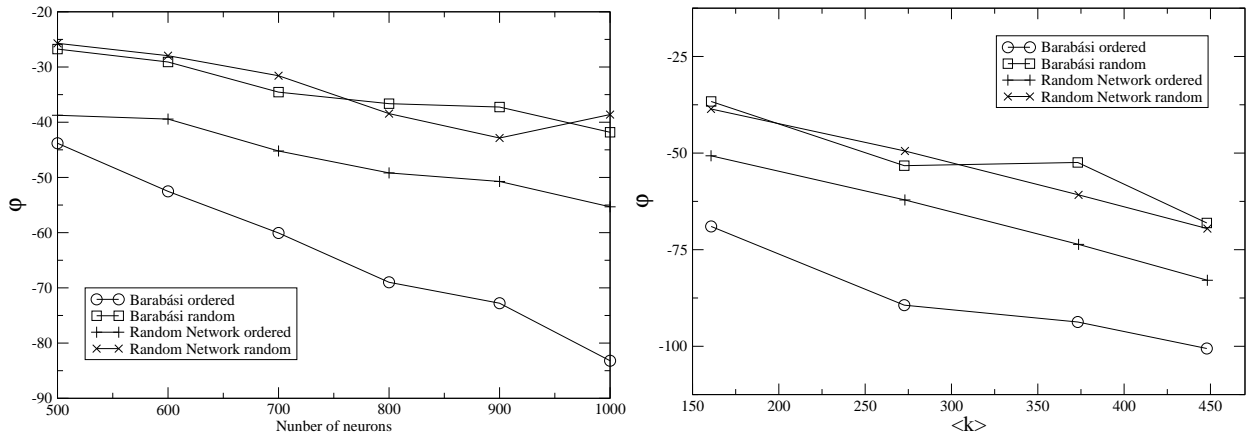

       \includegraphics[scale=0.35,clip]{figura4_I.eps}
       \includegraphics[scale=0.35,clip]{figura4_II.eps}
       \caption{The slope, defined in Figure~\ref{Rec_attack}, as a function of two important network parameters}. \label{theta}
\end{figure*}

The similar effect of random hub removal from both network models can be
better illustrated by defining a~\emph{recall index} as corresponding to the
number of patterns at which the respective overlap value reaches $0.8$.  We
also define the~\emph{attack rate} as the number of eliminated hubs divided by
the total number of neurons. Figure~\ref{Rec_attack} shows the degradation of
the recall index for each experiment as a function of the attack rate, showing
clearly the different behavior of a Barabasi network under orderly hub
elimination.  Yet, Figure \ref{theta}, shows the stability of such behavior
under variations of network parameters $N$ and $\left\langle k \right\rangle$.

These results support the intuitive expectation about hub importance for the
inner workings of the Hopfield memory model. Note nevertheless that the
comparisons between Random and Barab\'asi network were carried out regarding
only hub attack and, in this manner, it is not guaranteed that the same amount
of links or edges is eliminated from both networks. In what follows, we carry
out further experiments in order to evaluate the relative importance of edges
and hub removal.

\begin{figure*}
        \begin{center}
        \begin{tabular}{lr}
        A)\includegraphics[scale=0.2,clip,angle=-90]{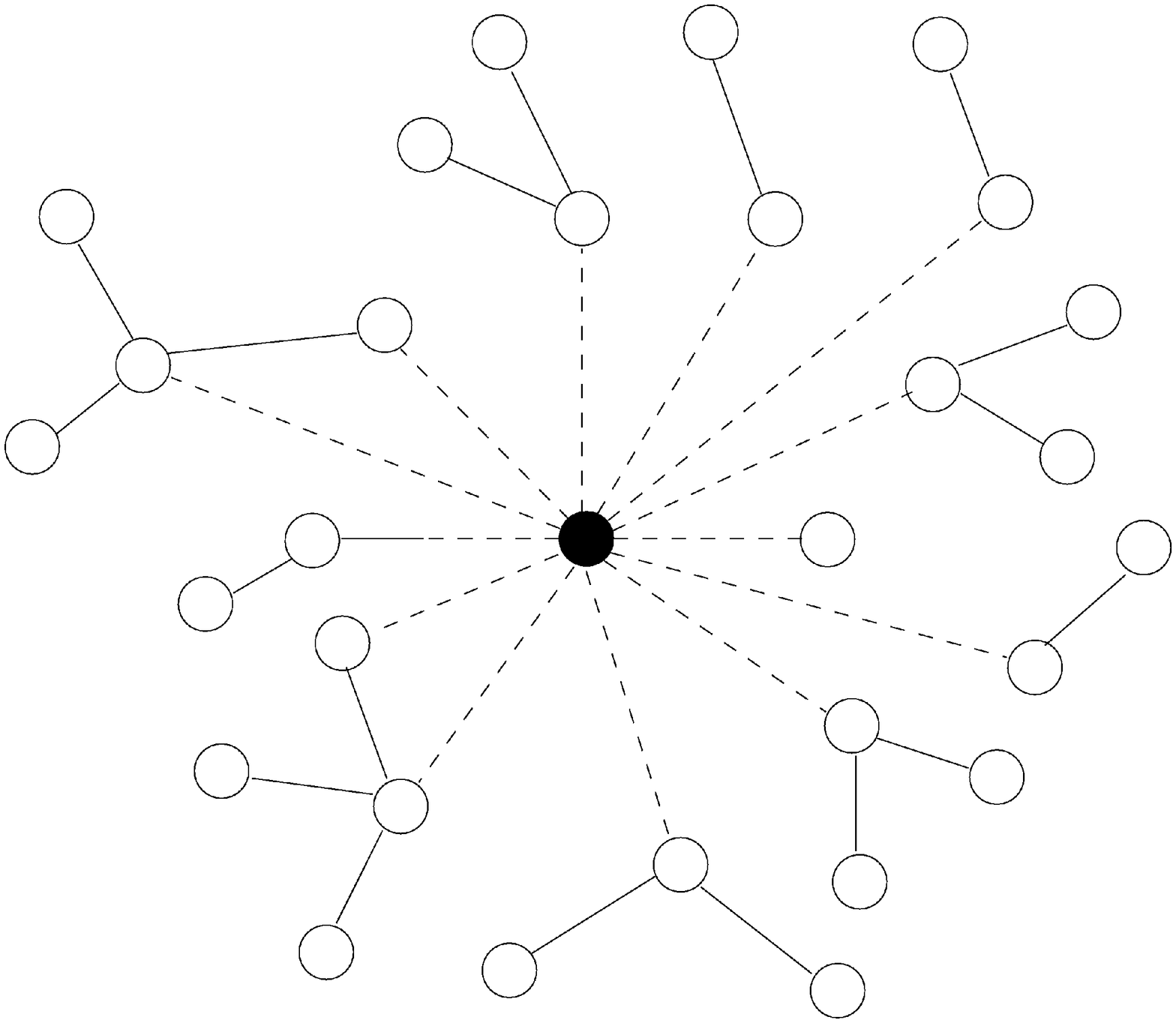}&
        B)\includegraphics[scale=0.2,clip,angle=-90]{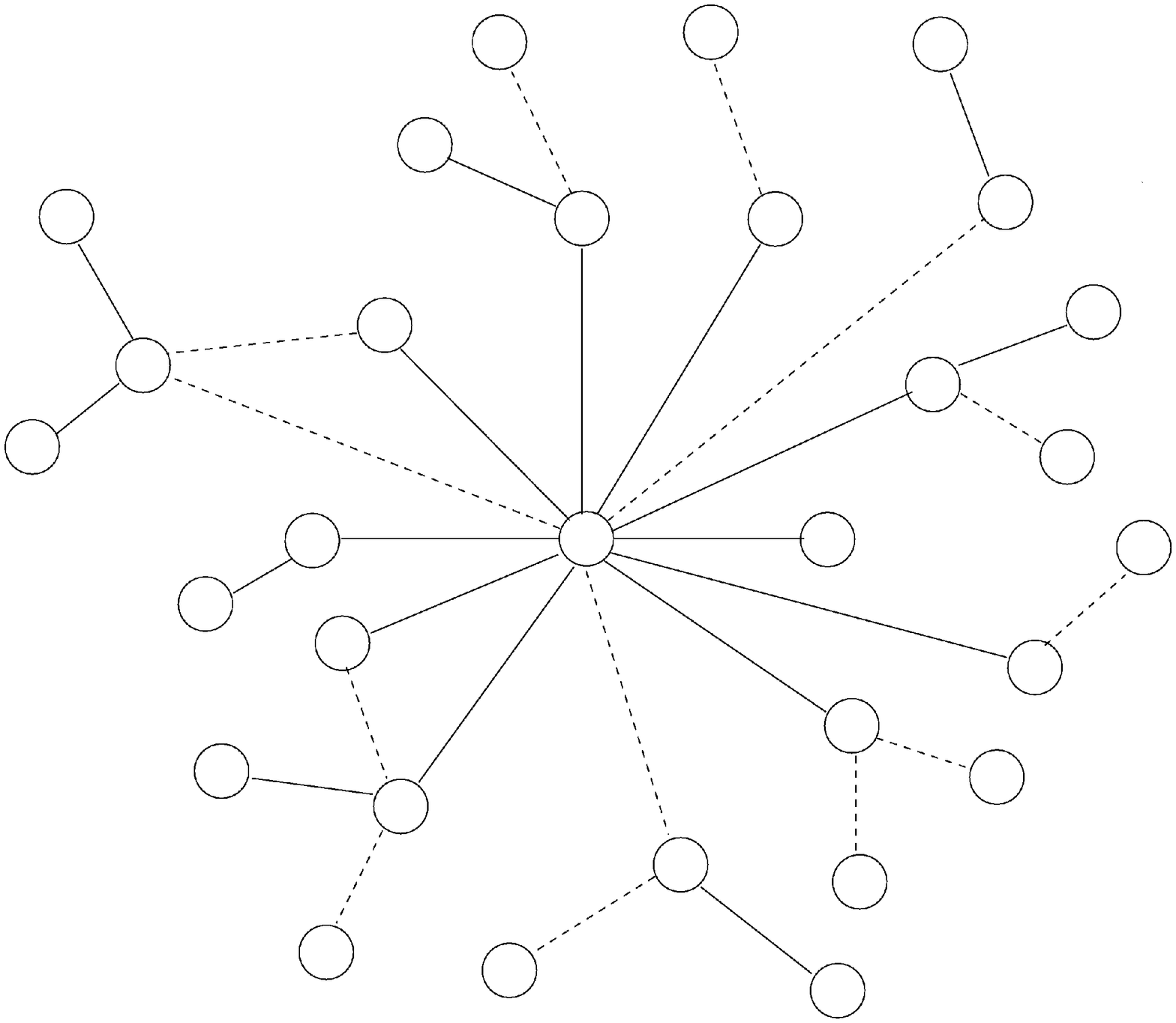}\\\\
        C)\includegraphics[scale=0.2,clip,angle=-90]{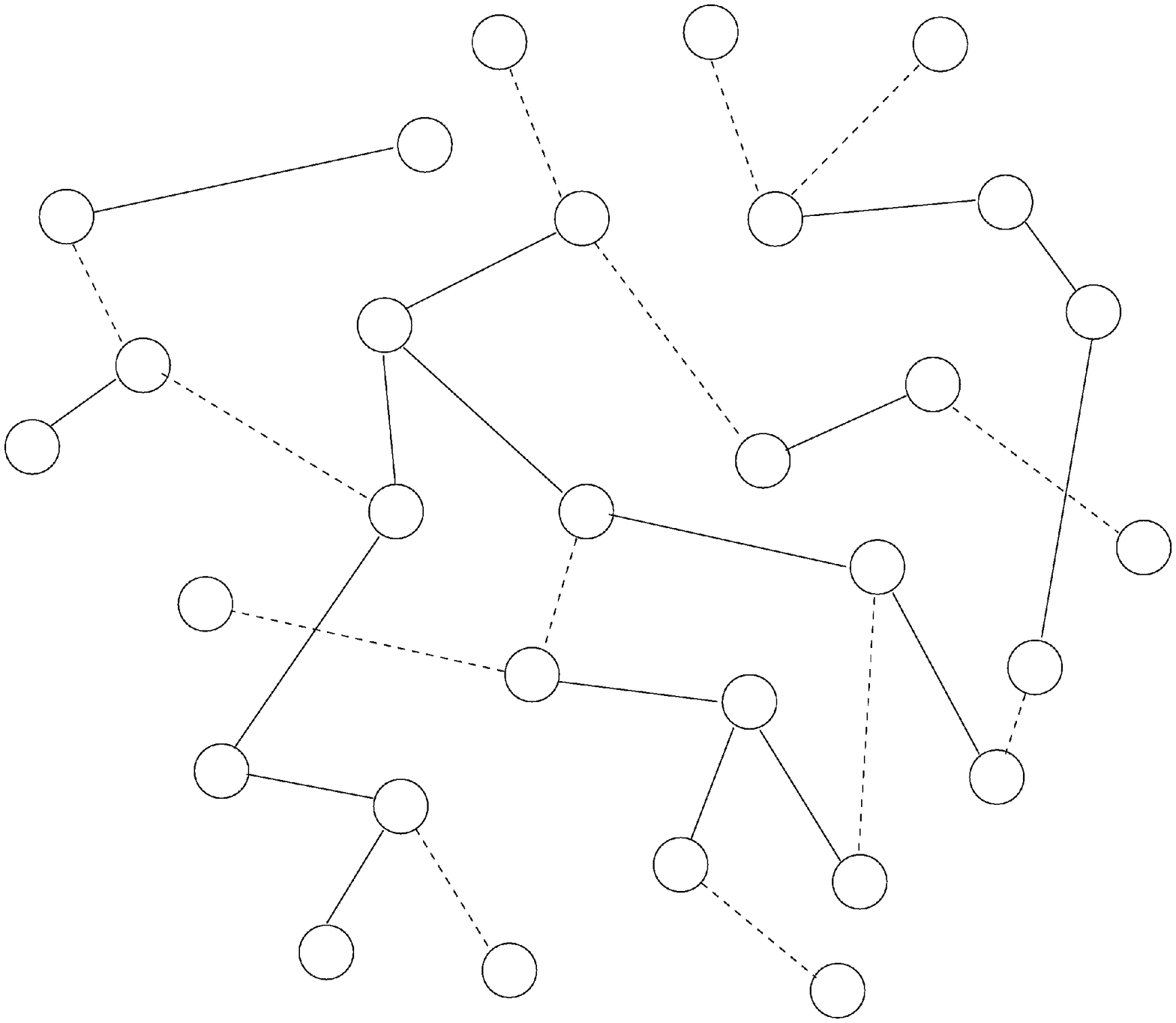}&
        D)\includegraphics[scale=0.2,clip,angle=-90]{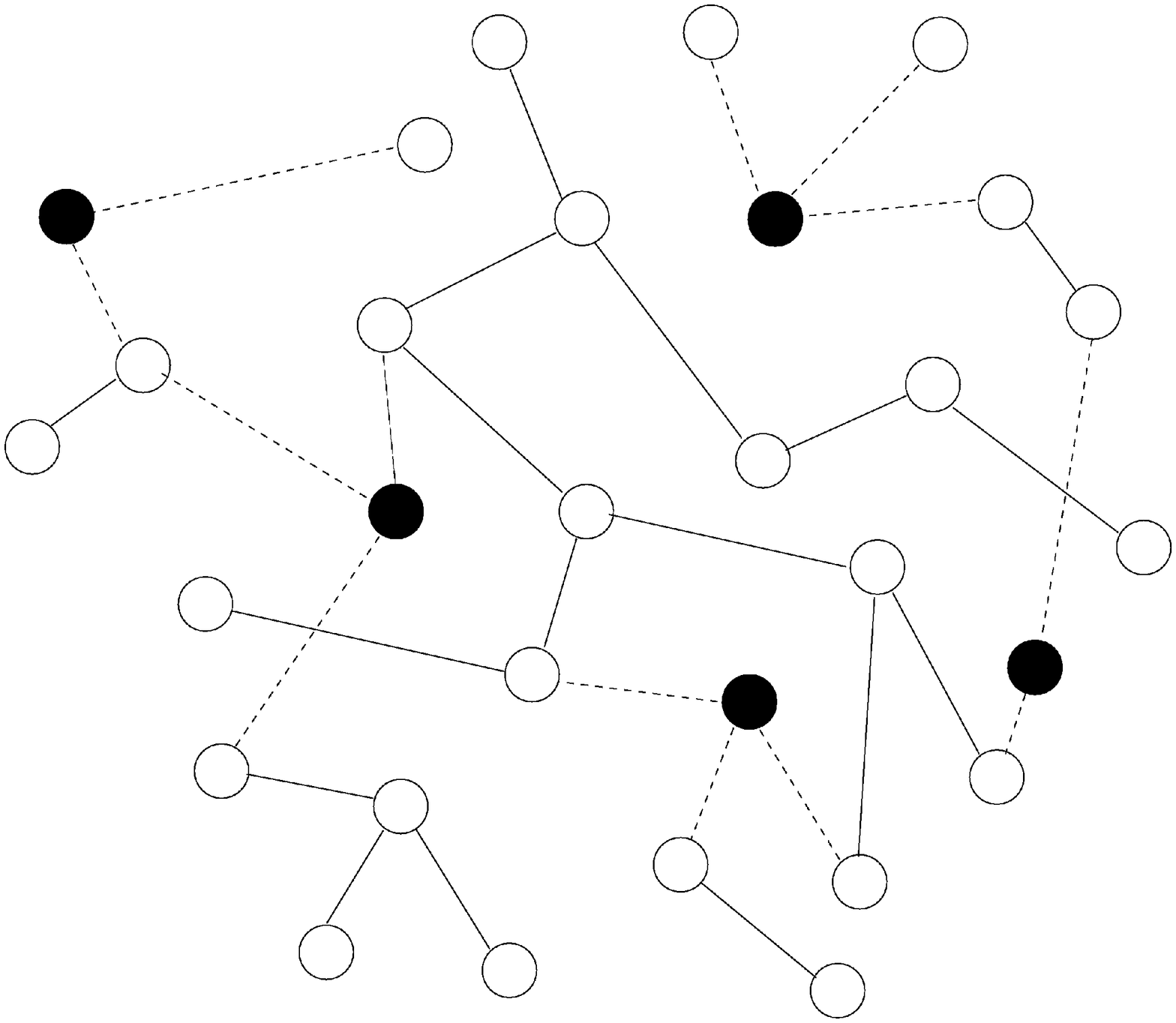}
        \end{tabular}
        \caption{A - hub attack to Barab\'asi network (black circle), eliminating
$e=13$ edges (dashed line),
                B - edge attack to the same Barab\'asi network  eliminating
$e=13$ edges at random,
                C - edge attack to Random network eliminating $13$ edges at
random,
                D - node attack to Random network eliminating nodes until the
amount of eliminated
                edges reaches $e=13$, that gives a total of $5$ nodes.}
       \label{net_graph}
       \end{center}  
\end{figure*}

\begin{figure*}
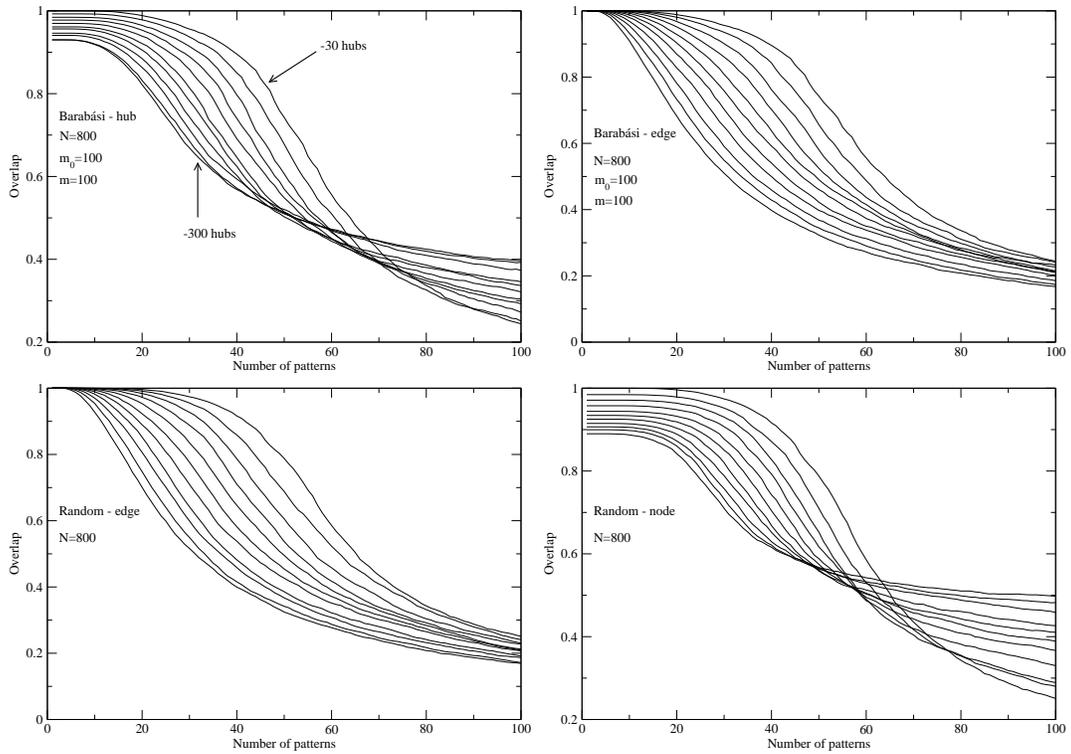

       \begin{center}
       \begin{tabular}{lr}
       \includegraphics[scale=0.3,clip]{figura6_I.eps}&
       \includegraphics[scale=0.3,clip]{figura6_II.eps}\\
       \includegraphics[scale=0.3,clip]{figura6_III.eps}&
       \includegraphics[scale=0.3,clip]{figura6_IV.eps}
       \end{tabular}
       \caption{The corresponding overlap curves for the situations described in Figure \ref{net_graph}}
       \label{kill_edge}
       \end{center}
\end{figure*}

\begin{figure*}
\centering
\includegraphics[scale=0.4,clip]{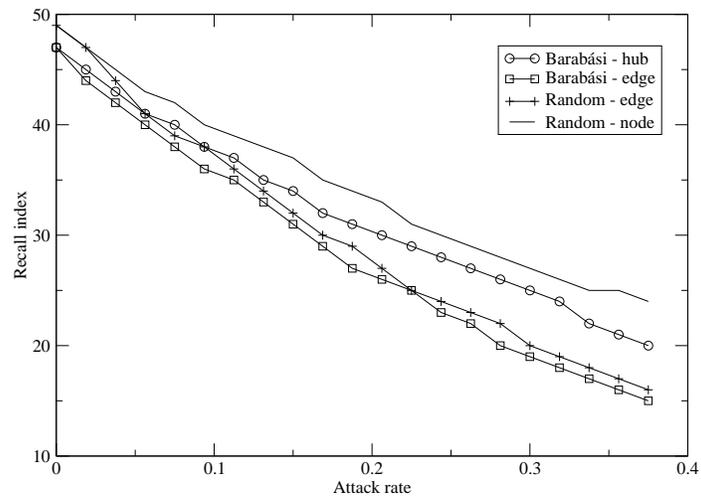}
\caption{Recall index as a function of the attack rate for edge attack to both
  types of networks}
\label{edg_attack}
\end{figure*}

We consider three experiments ensuring that the number of edges eliminated
from each networks is always the same. These experiments are illustrated in
Figure~\ref{net_graph}. In the first step, the hubs in a Barab\'asi network
are attacked, as indicated in Figure \ref{net_graph}{A}. The black circle
represents a hub and the dashed line its $e$ eliminated edges. From an
identical Barab\'asi network we then eliminate $e=6$ edges at random, as
shown in Figure~\ref{net_graph}{B}. In our experiments with large networks
this procedure is repeated until $300$ hubs are eliminated. The overlap curve
for both networks obtained at each step, after another $30$ hubs were
eliminated, is shown in Figure~\ref{kill_edge}(A and B). For an equivalent
Random network, the $e$ edges are eliminated either at random or by removing
nodes at random until the total amount of eliminated edges equals $e$, as
shown in Figure \ref{net_graph} C and D, respectively.  In the actual
experiments, overlaps curves are again produced for each set of $30$
eliminated hubs.  The results for this experiments are showed in the Figure
\ref{kill_edge} C and D, respectively. Figure~\ref{edg_attack} shows the
performance of the networks under attack, quantified by the recall index, in
terms of the attack ratio.

\section{Discussion and Conclusion}

Although largely overlooked by initial studies in artificial neuronal
networks, the connectivity between neurons plays a decisive role in
constraining and even defining the dynamical properties of the neuronal
systems, especially the potential of memory recall.  The results obtained in
our investigation substantiate further such a phenomenon, with special
attention focused on the importance of hubs to the overall memory recall, as
quantified by the overlap index.  The main interesting conclusions identified
from our experiments are listed and discussed in the following:

\begin{itemize}
\item \emph{Effect of connectivity dilution:} As expected, all types of edge
  and node dilutions act in order to reduce the recall in the Hopfield model,
  for both Barab\'asi and Random networks. See Figure~\ref{diff_mean_k}.
  
\item \emph{Ordered hub attack:} The removal of hubs, starting with the most
  connected ones, from Barab\'asi networks tends to cause more severe loss of
  recall performance than when removing hubs from random networks. This
  conclusion is implied by the fact that the curve for ordered attack to hubs
  in Barab\'asi networks, see Figure~\ref{Rec_attack}, correspond to the
  strongest recall deterioration. This reflects the relative importance of the
  hubs in Barab\'asi-Albert neuronal networks, as a consequence of the fact
  that hubs in such nets tend to have higher node degree (and therefore more
  edges) than hubs in random networks with similar average degree.
  
\item \emph{Random node attack:} Random attack to both types of networks did
  not tend to produce different performance reduction, as indicated by the two
  entangled curves representing random attack to both types of networks in
  Figure~\ref{edg_attack}.
  
\item\emph{Localized effect of hub removal:} A particulalry interesting point
  in the obtained results, for both types of network models, regards the fact
  that the random removal of the same quantity of edges as in a large hub from
  the same network has a stronger effect on the recall performance than
  removing that hub. This property is supported by the fact that the curves,
  in Figure~\ref{edg_attack}, for hub (node) attack are higher than the
  respective edge attack curves. This is a direct consequence of the fact that
  the removal of the hub implies more localized connectivity deterioration,
  affecting predominantly the single bit of the recovered pattern which is
  directly related to the hub.

\end{itemize}

All in all, it has become clear that hub and edge attack, even when involving
the same number of links, have quite distinct effects on the memory recall of
Hopfield networks running on random and Barab\'asi-Albert networks.

Future works related to the currently reported investigation should
consider the effect of edge and hub attack on other network topologies
such as small-world~\cite{Albert_Barab:2002} and
Sznajd~\cite{Sznajd:complex}.  It
would also be interesting to extend such experimental analyses to
artificial neuronal network models such as the
perceptron~\cite{haykin:1999} and self-organizing
maps~\cite{kohonen:2001}.

\acknowledgments
The authors are financially supported by grants FAPESP(02/02504-01 and 99/12765-2) and CNPQ(308231/03-1).

\nocite{*}
\bibliographystyle{plain}
\bibliography{Hopfield}

\end{document}